# Band structure and energy level alignment of chiral graphene nanoribbons on silver surfaces

Martina Corso,*,a,b Rodrigo E. Menchón,b Ignacio Piquero-Zulaica,a, † Manuel Vilas-Varela,c J. Enrique Ortega,a,b,d Diego Peña,c Aran Garcia-Lekue,*,b,e and Dimas G. de Oteyza*,a,b,e

**Chiral graphene nanoribbons are extremely interesting structures due to their low bandgaps and potential development of spin-polarized edge states. Here, we study their band structure on low work function silver surfaces and assess the effect of charge transfer on their properties.**

Carbon-based nanostructures can display exceptionally varied properties depending on their precise bonding structure. This includes graphene nanoribbons (GNRs),[1–3] in which a graphene lattice is confined to narrow one-dimensional stripes. GNRs with armchair-oriented edges display a semiconducting band structure. In contrast, zigzag and even chiral GNRs are quasi-metallic and develop spin-polarized edge states,[2–5] unless they are exceedingly narrow. In this case, the edge states from either side hybridize with one another, which quenches the spin-polarization and confers the ribbons a conventional semiconducting band structure.[6,7]

For ribbons with a (3,1) chiral vector, the minimum width required to maintain the quasi-metallic behaviour comprises six carbon zigzag lines from side to side.[6] This theoretical prediction has been recently confirmed experimentally by synthesizing and spectroscopically characterizing (3,1) chiral GNRs of varying width on Au(111).[8] However, these ribbons, as it occurs also with purely zigzag-edged GNRs,[9] or with other GNRs featuring low energy states associated to periodic zigzag edge segments,[10–12] have been synthesized and characterized to date only on Au(111).

To investigate the effect of different substrates with lower workfunction on the ribbon's electronic properties, here we synthesize six zigzag lines wide (3,1) chiral GNRs ((3,1,6)-chGNRs, Fig. 1a) on a curved Ag crystal[13] that spans up to ±15 degrees of vicinal angle α to either side with respect to the central (111) surface orientation (Fig. 1b). The synthesis is successful over the entire crystal, but the different types of steps on each side of the sample have a disparate effect on the ribbon's preferred azimuthal alignment. This provides us with an ideal sample on which to study the band dispersion by angle resolved photoemission (ARPES), both along and perpendicular to the ribbon's longitudinal axis.

The reactant we use is 2′,6′-dibromo-9,9′:10′,9″-teranthracene (DBTA, Fig. 1a), synthesized as described in the supplementary information.[8] It transforms into (3,1,6)-chGNRs following a two-step process that consists of thermally activated Ullmann coupling and cyclodehydrogenation (CDH, Fig. 1a).[8] The substrate is a Ag single crystal curved around the [1,1,1] axis, with the (111) surface plane at the crystal's central area as displayed in Fig. 1b.[13] The stepped surfaces towards either side thus share the same (111) terrace structure (of varying width $d$ depending on the vicinal angle). However, the steps display non-equivalent facets, namely {100} facets on the left hand side and {111} facets on the right hand side (Fig. 1b).[13]

Figure 2 shows representative images of the sample after depositing nearly a full monolayer of precursor molecules (DBTA) and stepwise annealing the crystal to 180 ºC for 10 minutes and to 315 ºC for 1 minute, to drive the subsequent activation of polymerization and cyclodehydrogenation. Notably, the resulting GNRs display three distinct arrangements depending on the region of the Ag crystal.

On the left hand side of the crystal the GNRs are found preferentially aligned parallel to the steps direction (Fig. 2a). In fact, in addition to the uniaxially aligned ribbons on top of the flat terraces, the ribbons display a particular affinity to the steps, adsorbing in a tilted configuration with either side of the GNR on each of the two neighbouring terraces (Fig. 2a). Both of these findings were expected, since stepped surfaces have been used often for alignment purposes[7,14–18] and many molecules, including GNRs,[16,19] are known to display particular affinity for adsorption on the undercoordinated and thus more reactive step atoms.

In the central part of the crystal, displaying ample flat terraces, the ribbons adsorb with multiple azimuthal orientations, as expected from the six-fold symmetry of the surface (Fig. 2b). On the right hand side of the crystal, the GNRs display again uniaxially aligned ribbons. However, the ribbons surprisingly align perpendicular rather than parallel to the steps' direction and extend over multiple terraces (Fig. 2c). Taking into account that the terraces on the right and left hand side of the crystal are identical, the difference in the preferential alignment must necessarily have its origin on the nature of the steps, which are formed by {100} and {111} facets, respectively. The specific interactions that cause this striking difference are beyond the scope of this work, but the resulting sample is ideal to probe the band dispersion parallel and perpendicular to the ribbon's axis by ARPES. We have performed ARPES measurements that record the dispersion parallel to the [-1,1,0] substrate direction, which coincides with the step direction both on the left and right hand side of the crystal. For the GNR bands, it corresponds to the dispersion along the longitudinal (Fig. 3(a)) and transverse direction of the ribbons (Fig. 3(c)), respectively, and to contributions from both in the central flat part of the crystal (Fig. 3(b)). The raw data, along with the reference measurements on the clean crystal are displayed in Fig. S1. As previously observed with narrower (3,1,4)-chGNRs,[7] the band

a. Centro de Fisica de Materiales (MPC), CSIC-UPV/EHU, 20018 San Sebastián, Spain.
b. Donostia International Physics Center (DIPC), 20018 San Sebastián, Spain.
c. Centro Singular de Investigación en Química Biolóxica e Materiais Moleculares (CiQUS) and Departamento de Química Orgánica, Universidade de Santiago de Compostela, Santiago de Compostela, Spain.
d. Universidad del País Vasco, Dpto. Física Aplicada I, 20018 San Sebastián, Spain
e. Ikerbasque, Basque Foundation for Science, Bilbao, Spain
† Current address: Physics Department E20, Technical University of Munich, 85748 Garching, Germany.
* martina.corso@ehu.eus; wmbgalea@ehu.eus; d_g_oteyza@ehu.eus

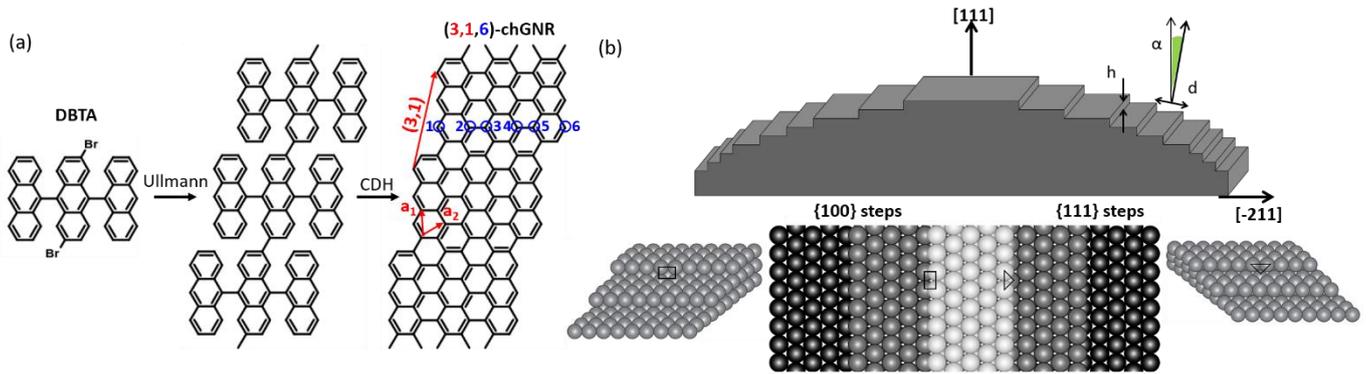

**Fig. 1**. (a) Reactant (DBTA) and reaction scheme towards the (3,1,6)-chGNR structure, displaying a (3,1) chiral vector marked in red and six atoms across its width marked in blue. (b) Schematic description of the Ag curved crystal where $d$ corresponds to the terrace width, $\alpha$ to the vicinal angle from the [111] direction and $h$ to the monoatomic step height. The steps at the left and right sides of the crystal display {100}-oriented and {111}-oriented microfacets, respectively.

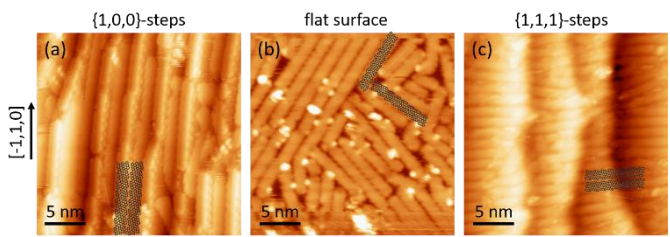

**Fig. 2**. Representative STM images of the sample after GNR synthesis in regions with {100}-steps (a), on the flat (111) surface in the central crystal region (b), and with {111}-steps (c). The [-1,1,0] direction that coincides with the steps direction is shown on the left. Segments of two GNR structures are superimposed on each of the images as a guide to the eye.

dispersion along the longitudinal ribbon direction (Fig. 3a) is hardly recognizable in the first Brillouin zone, starts becoming visible in the second and appears most intense in the third Brillouin zone (centred around 1.4 Å$^{-1}$). Indeed, in the third Brillouin zone not only the valence band but also following bands are observed with remarkable clarity, allowing for a direct comparison with the band structure predicted by DFT calculations for free-standing ribbons. As pictured in Fig. 3a with the calculated bands superimposed on the ARPES data, there is an excellent match between experiment and theory. Such good match, however, requires shifting the charge neutrality point (CNP) by -0.52 eV.

This shift implies charge transfer at the GNR/silver interface. In contrast to Au, on which the GNRs show a clear tendency to become $p$-doped,[8,20] the substantially lower work function of silver (e.g. 4.6 eV as compared to 5.4 eV for the (111) surfaces of Ag and Au,[21] respectively) favours the opposite electron transfer from surface to GNR. For the ARPES characterization, whereby only filled states are accessed, this has the advantage that also the conduction band can be probed. To which extent the conduction band becomes accessible (populated) is quantitatively related to the charge transfer, taking into consideration that each band hosts two electrons per unit cell. The measurements in Fig. 3a display 51% of the conduction band below the Fermi level, from where we can conclude that approximately one electron per GNR unit cell is transferred from the silver surface to the (3,1,6)-chGNRs, in qualitative agreement with the 1.3 electrons required to shift the CNP by 0.52 eV according to DFT calculations.

However, the amount of charge transfer shows variations across the curved silver surface. Figure 3c displays the dispersion along the transverse ribbon's axis. Along this direction, the electronic states do not show any notable dispersion and appear as flat bands. This implies a negligible overlap of the wavefunctions of electronic states in neighbouring ribbons. The extent to which the conduction band is populated cannot be inferred from these data as clearly as before. Yet, the flat band associated with the charge neutrality point appears at a similar energy as in Fig. 3(a), and hence the charge transfer can be concluded to be comparable. The situation is slightly different in the central region of the crystal (Fig. 3(b)). There, the ribbons display multiple azimuthal orientations, each of them contributing to the convoluted ARPES signal. The overall dispersion can thus be recognized less clearly, although the general appearance can be ascribed to a washed-out convolution of Fig. 3a and 3b. However, the CNP appears about 0.16 eV higher in energy. As a result, only 31% of the conduction band is populated, which in turn implies a charge transfer of only about 2/3 of an electron per unit cell. The modulation of the CNP as a function of the vicinal angle is displayed in Fig. 3d (see the associated data in Fig. S2) and is ascribed to the lower work function in the stepped regions as compared to the compact flat surface.[7,16,22] Figure 3d also displays the charge transfer required to shift the CNP to the measured energies, according to DFT calculations, and underlines the importance of local work function variations for influencing the electronic properties in weakly interacting metal-organic interfaces.[23,24]

Indeed, theoretical calculations on ribbons with different doping levels reveal important implications for their properties.[25,26] Whereas in the absence of spin-polarization (3,1,6)-chGNRs display a quasi-metallic band structure,[8] allowing for spin-polarization results in a 16 meV more favourable ground state that includes an increased bandgap and edge states with antiferromagnetically oriented magnetization (Fig. 4a). However, charging the system with 1.3 extra electrons per unit cell shifts the CNP by -0.52 eV, quenches the magnetization and recovers the quasi-metallic band structure with no spin-polarization (Fig. 4b).

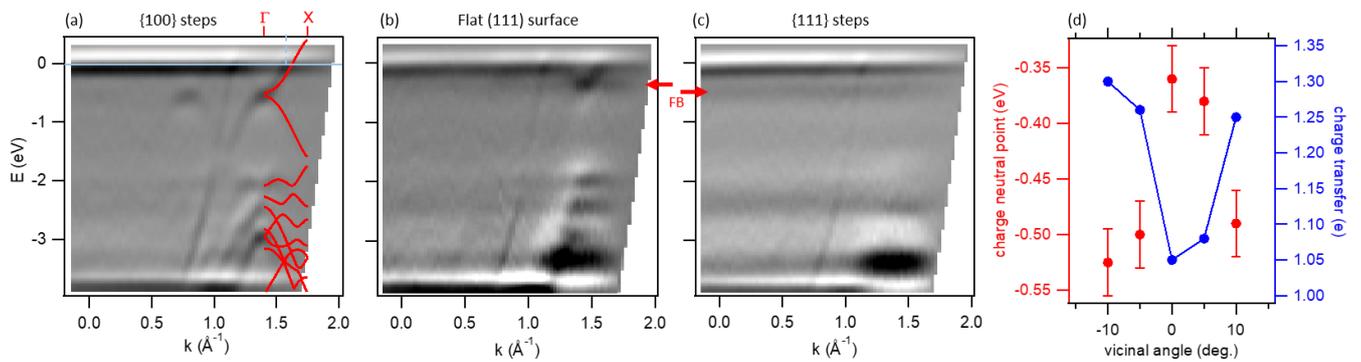

**Fig. 3**. ARPES data displaying the dispersion along the [-1,1,0] direction of the curved Ag crystal on the stepped regions with {100} step facets (a, α≈-10°) on the central flat region (b, α≈0°) and on the stepped region with {111} step facets (c, α≈10°). The calculated band structure for free-standing GNRs after shifting the charge neutral point to -0.52 eV is superimposed on the third Brillouin zone of panel (a). The horizontal light blue solid line marks the Fermi energy and the vertical light blue dashed line marks its crossing point with the CB. The red arrows in panels (b) and (c) mark the flat band (FB) at the charge neutral point. (d) Measured CNP as a function of the vicinal angle and calculated charge transfer to reach such interface band alignment, according to DFT.

In an attempt to quantify the necessary charge to quench the edge state magnetization, we have performed additional calculations gradually modifying the GNR doping level. As depicted in Fig. 4(c), a charge transfer of only 0.3 electrons per unit cell is already sufficient to fully prevent any magnetism in this kind of ribbons. Although the exact value may vary for nanoribbons of different width or chirality, this is a key finding to keep in mind for the design of potential devices aiming at the exploitation of the magnetic edge states of GNRs.

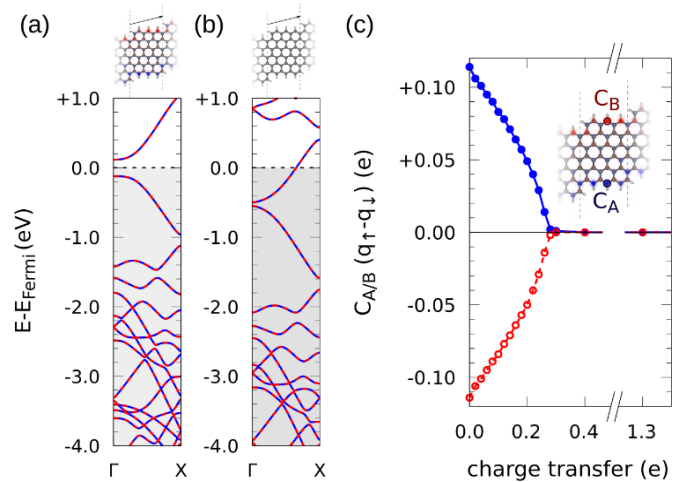

**Fig. 4**. Spin density (top) and calculated band structure (bottom) for neutral (3,1,6)-chGNRs (a), and upon charge transfer of 1.3 electrons per unit cell (b). (c) Spin-polarized electron density at the marked carbon atoms at the ribbon's edges as a function of charge transfer.

In our experiment, the charge transfer throughout the whole crystal is such that it fully quenches the magnetization. However, higher work function materials may instead provide an energy level alignment that maintains the intrinsic edge state spin polarization, and the smoothly varying work function in curved crystals as a function of the vicinal angle may help in its fine adjustment.[22]

In conclusion, we have synthesized chiral graphene nanoribbons on a curved silver crystal. Depending on the types of underlying steps, the ribbons grow along different orientations, while the step density modulates the substrate work function. This allows probing the energy level alignment and band dispersion in different configurations. By theoretical calculations we have further analyzed the consequences for the ribbon's edge state magnetization, which is fully quenched for as low amounts of charge transfer as 0.3 electrons per GNR unit cell.

We acknowledge financial support from the Agencia Estatal de Investigación (grant nos. PID2019-107338RB-C62, PID2019-107338RB-C63, PID2019-107338RB-C66 and MAT-2017-88374-P), the European Union's Horizon 2020 research and innovation program (grant no. 863098), the Xunta de Galicia (Centro Singular de Investigación de Galicia, 2019-2022, grant no. ED431G2019/03), the European Regional Development Fund and the Basque Government (Grant IT-1255-19).

## Notes and references

# Supporting Information:

# Band structure and energy level alignment of chiral graphene nanoribbons on silver surfaces

Martina Corso,*[a,b] Rodrigo E. Menchón,[b] Ignacio Piquero-Zulaica, [a, †] Manuel Vilas-Varela,[c] J. Enrique Ortega,[a,b,d] Diego Peña,[c] Aran Garcia-Lekue,*[b,e] and Dimas G. de Oteyza*[a,b,e]

[a.] *Centro de Fisica de Materiales (MPC), CSIC-UPV/EHU, 20018 San Sebastián, Spain.*
[b.] *Donostia International Physics Center (DIPC), 20018 San Sebastián, Spain.*
[c.] *Centro Singular de Investigación en Química Biolóxica e Materiais Moleculares (CiQUS) and Departamento de Química Orgánica, Universidade de Santiago de Compostela, Santiago de Compostela, Spain.*
[d.] *Universidad del País Vasco, Dpto. Física Aplicada I, 20018 San Sebastián, Spain*
[e.] *Ikerbasque, Basque Foundation for Science, Bilbao, Spain*
† Current address: Physics Department E20, Technical University of Munich, 85748 Garching, Germany.
* martina.corso@ehu.eus; wmbgalea@ehu.eus; d_g_oteyza@ehu.eus

**Synthesis of 2',6'-dibromo-9,9':10',9''-teranthracene (DBTA)**

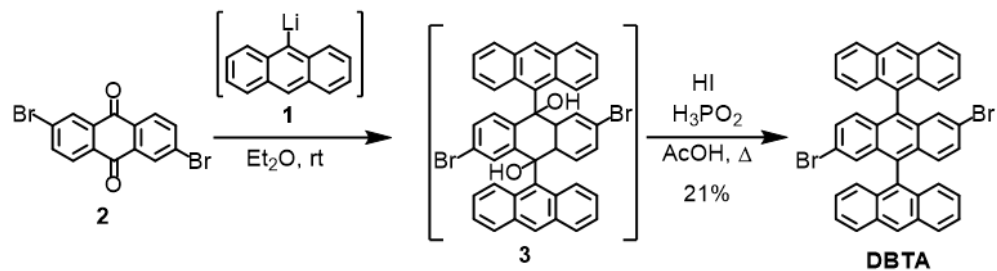

As previously reported,[1] over a solution of organolithium **1** (30 mL, 1.37 mmol, 0.05 M in THF), a suspension of anthraquinone **2** (200 mg, 0.55 mmol) in THF (10 mL) was added at r.t. The resulting mixture was stirred for 20 h at r.t. Then, AcOH (1.0 mL) was added and volatiles were removed under reduced pressure. The residue (compound **3**) was dissolved in AcOH (20 mL), and a mixture of $H_3PO_2$ (5.00 mL, 45.5. mmol, 50%) and HI (0.80 mL, 6.08 mmol, 57%) were added and the resulting mixture was heated at 80 ºC for 2 h. After cooling to r.t., the mixture was diluted with $CH_2Cl_2$ (60 mL) and $H_2O$ (60 mL), the phases were separated and the organic layer was washed with $NaHCO_3$ (saturated aqueous solution, 3x30 mL). The organic phase was dried over anhydrous $Na_2SO_4$, filtered and evaporated. The residue was purified by column chromatography ($SiO_2$; Hex:$CH_2Cl_2$ 4:1 to 7:3), affording **DBTA** (77 mg, 21%) as a yellow solid.

**$^1$H NMR** (363 K, 500 MHz, $C_2D_2Cl_4$) δ: 8.72 (s, 2H), 8.18 (d, *J* = 8.5 Hz, 4H), 7.50 (dd, *J* = 8.9, 5.9 Hz, 4H), 7.34 (d, *J* = 2.0 Hz, 2H), 7.32 – 7.26 (m, 4H), 7.21 (d, *J* = 8.7 Hz, 4H), 7.11 (dd, *J* = 9.3, 2.0 Hz, 2H), 7.01 (d, *J* = 9.3 Hz, 2H) ppm. **$^{13}$C NMR-DEPT** (363 K, 125 MHz, $C_2D_2Cl_4$) δ: 134.2 (2C), 132.8 (4C), 131.9 (6C), 131.7 (2C), 130.7 (2C), 130.6 (2C), 130.2 (2CH), 129.4 (2CH), 129.1 (6CH), 128.3 (2CH), 126.8 (4CH), 126.6 (4CH), 125.8 (4CH) ppm. **MS (EI)** m/z (%): 688 (M+, 100), 522 (9), 344 (22), 261 (22), 255 (28). **HRMS**: $C_{42}H_{24}Br_2$; calculated: 686.0245, found: 686.0224.

*1H and 13C NMR spectra of DBTA:*

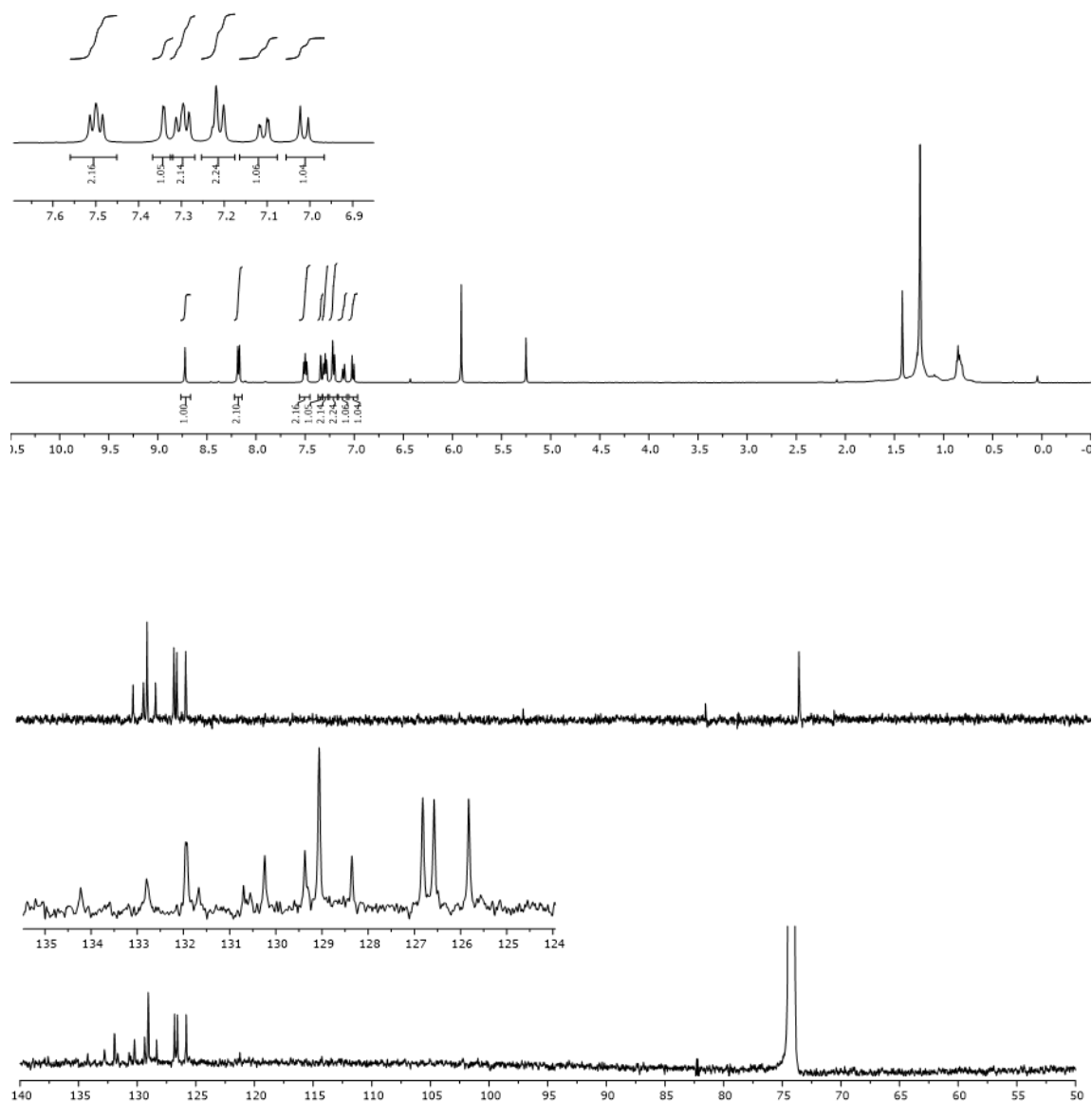

**Experimental surface science methods**
The employed curved silver crystal was prepared with standard sputtering/annealing parameters (E=1000 eV / T=370 ºC). The reactant molecules were sublimed from a home-made Knudsen cell heated to a temperature of around 265 ºC at a rate of 0.06 ML/min as controlled with a calibrated quartz crystal microbalance. The sample was subsequently annealed to 180 ºC and 315ºC for 10 min and 1 min, respectively, to separately activate the polymerization and cyclodehydrogenation steps. The sample was first analyzed with STM and subsequently transferred to the ARPES chanber without breaking the vacuum. The STM images were acquired with a commercial Omicron VT-STM and processed with the WSXM software.[2] ARPES measurements were obtained with a high-intensity monochromatic source (21.2 eV) and a high-resolution display-type hemispherical electron analyzer (Phoibos150). The vertically aligned manipulator and analyzer slit were perpendicular to the horizontally aligned step direction of the curved crystal, allowing measurements over a wide band dispersion range parallel to the steps by sample rotation (polar scans by manipulator rotation). The sample temperature during the ARPES experiments was approximately 150 K.

**Calculations**
First-principles electronic structure calculations were performed using DFT as implemented in the SIESTA software package.[3,4] The van der Waals density functional by Dion et al.[5] with the modified exchange correlation by Klimeš, Bowler and Michaelides[6] was used. The valence electrons were described by a double-ζ plus polarization (DZP) basis set with the orbital radii defined using a 54 meV energy shift,[4] while the core electrons were described using norm-conserving Trouillers-Martins pseudopotentials.[7] For integrations in real space[4] an energy cutoff of 300 Ry was used. The smearing of the electronic occupations was defined by an electronic temperature of 300 K with a Fermi-Dirac distribution. The self-consistency cycles were stopped when variations on the elements of the density matrix were less than $10^{-4}$ eV and less than $10^{-4}$ eV for the Hamiltonian matrix elements. In order to avoid interactions with periodic images from neighboring cells, systems were calculated within a simulation cell where at least 50 Å of vacuum space was considered. Variable cell relaxations and geometry optimizations were performed using the conjugate gradient method using a force tolerance equal to 10 meV/Å and 0.2 GPa as a stress tolerance. A 101 k-point mesh along the GNRs' periodic direction was used.

**Supplementary data**

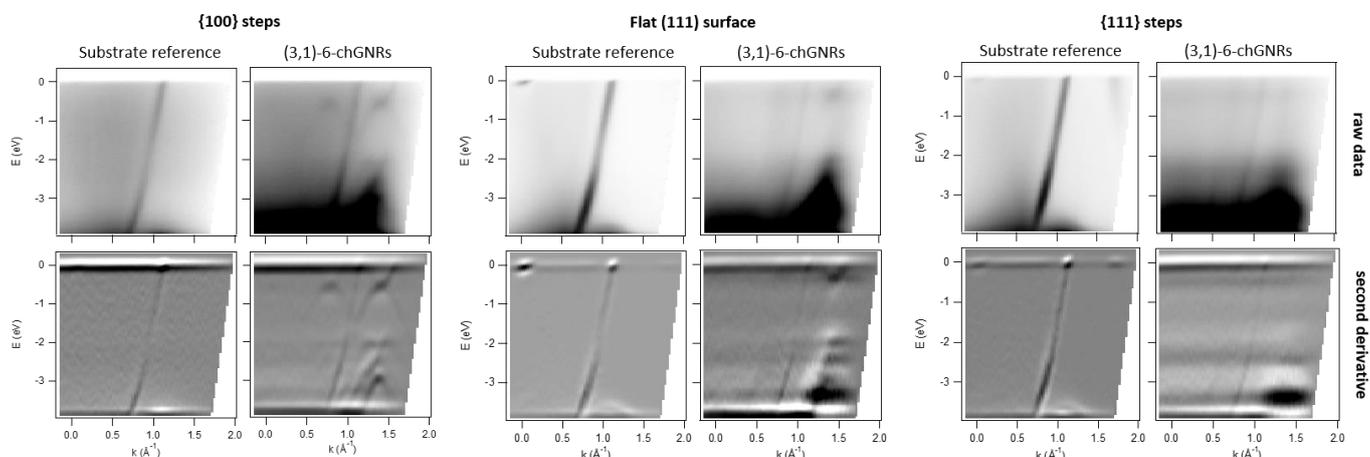

**Fig. S1**. ARPES raw data and their second derivative for three representative regions of the GNR-covered curved Ag crystal surface (characterized by periodic {100} steps, flat (111) surface and periodic {111} steps), along with the reference map on the clean substrate. The displayed measurements are obtained at vicinal angles of approximately -10°, 0 and 10°.

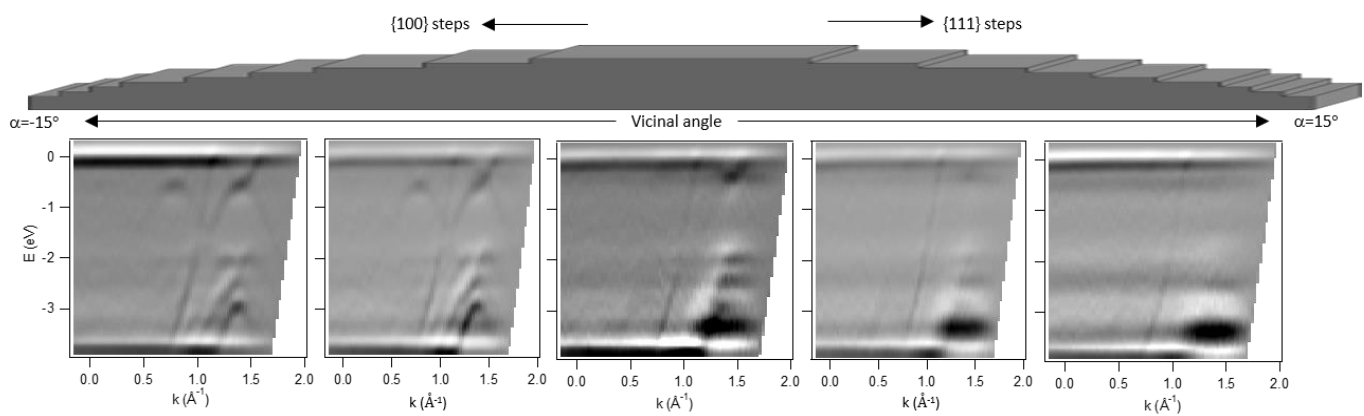

**Fig. S2**. ARPES data obtained at five regularly spaced positions across the curved silver surface at vicinal angles of approximately -10°, -5°, 0, 5° and 10°.